\documentclass[aps,prb,twocolumn,superscriptaddress,showpacs]{revtex4}
\usepackage{graphicx}
\usepackage{dcolumn} 
\usepackage{bm}      
\usepackage{amsmath,amssymb,times}
\usepackage{color}

\begin{document}

\title{Magneto-electric effect in NdCrTiO$_{5}$}

\author{J.~Hwang}
\affiliation{Department of Physics, Florida State University,
Tallahassee, FL 32306-3016, USA}
\affiliation{National High Magnetic
Field Laboratory, Florida State University, Tallahassee, FL
32310-3706, USA}

\author{E.~S.~Choi}
\affiliation{National High Magnetic
Field Laboratory, Florida State University, Tallahassee, FL
32310-3706, USA}

\author{H.~D.~Zhou}
\affiliation{National High Magnetic
Field Laboratory, Florida State University, Tallahassee, FL
32310-3706, USA}

\author{J.~Lu}
\affiliation{National High Magnetic
Field Laboratory, Florida State University, Tallahassee, FL
32310-3706, USA}

\author{P.~Schlottmann}
\affiliation{Department of Physics, Florida State University,
Tallahassee, FL 32306-3016, USA}
\affiliation{National High Magnetic
Field Laboratory, Florida State University, Tallahassee, FL
32310-3706, USA}

\date{\today}

\begin{abstract}
We have measured the dielectric constant and the pyroelectric current of orthorhombic (space group $Pbam$) NdCrTiO$_5$ polycrystalline samples. The dielectric constant and the pyroelectric current show features associated with ferroelectric transitions at the antiferromagnetic transition temperature ($T_{\text{N}}$ = 21 K). The effect of magnetic fields is to enhance the features almost linearly up to the maximum measured field (7 T) with a spontaneous polarization value of $\sim 3.5~\mu$C/m$^2$. Two possible scenarios, the linear magnetoelectric effect and multiferroicity (antiferromagnetism + ferroelectricity), are discussed as possible explanations for the observations.
\end{abstract}

\pacs{75.85.+t, 77.70.+a, 77.80.-e, 75.80.+q}
\maketitle

\section{Introduction}
Multiferroicity,\cite{Komskii,Eerenstein,Cheong,Ishihara} i.e. when two or more ferroic orders coexist in a single phase of a material, has attracted renewed great interest during the last a decade. Especially, multiferroics with magnetic and ferroelectric orders have been under intense investigation, both experimentally and theoretically, due to its possible application for memory devices. The common features of this type of multiferroics include dielectric anomalies and the emergence of spontaneous electric polarization at the magnetic ordering temperature, which are also strongly dependent on the strength and orientation of the magnetic field. In these materials, ferroelectricity is usually enabled via different types of exchange interactions which break the spatial inversion symmetry of a paraelectric phase. On the other hand, magnetoelectric (ME) materials show strong magnetic field dependent polarization and dielectric anomalies.\cite{Fiebig} The consideration of magnetic symmetry and the thermodynamic potential can give a great deal of information to understand the ME effect. For example, both broken spatial- and time-inversion symmetries are required for a linear ME effect due to the linear potential term ($EH$) in the thermodynamic potential, while the spatial-inversion symmetry can be conserved for a quadratic term ($EEH$).

Two mechanisms have been proposed as the macroscopic driving force for the symmetry breaking in a multiferroic phase transition, namely, the spin-current model \cite{Katsura,Sergienko} and magnetostriction\cite{goodenough}. Within the spin current model the symmetry breaking is due to the Dzyaloshinskii-Moriya (DM) interaction,\cite{Moriya,Dzyaloshinskii} where the antisymmetric exchange allows an electric polarization, ${\vec P}$, proportional to ${\vec e_{ij}} \times$ ${\vec S_i} \times {\vec S_j}$, where ${\vec e_{ij}}$ is the unit vector connecting two sites of spin ${\vec S_i}$ and ${\vec S_j}$. The spin current is more relevant to a system with spiral magnetic structure such as TbMnO$_3$\cite{Kenzelmann}, DyMnO$_3$\cite{Arima}, Ni$_3$V$_8$O$_8$\cite{Lawes}, CuFeO2\cite{Kimura_CuFeO2} and other compounds.\cite{Kimura_review} There are fewer examples to show multiferroicity with collinear magnetic structures. In Dy(Gd)FeO$_3$\cite{Tokunaga,gdfeo3} and Ca$_3$Co$_{2-x}$Mn$_x$O$_6$\cite{choi}, the DM term vanishes from symmetry and the magnetostriction mechanism was proposed to explain the multiferroicity. Another $Pbam$ multiferroic $R$Mn$_2$O$_5$ ($R$ = Y, Ho, Tb, Er) has more complicated spin structures with orderings of different commensurability. Its origin of ferroelectricity has been ascribed to either the spin current model\cite{KimuraH,Kim} or magnetostriction.\cite{chapon,Kim} Concerning the ME effect, regardless of the apparently different underlying mechanism, these types of multiferroics generally show large ME effect, which sometimes is linear $P_i$ = $\alpha_{ij} H_j$ or $M_i$ = $\alpha_{ij}E_j$ depending on the magnetic and crystallographic symmetries.

The crystalline and magnetic structure of powder samples of NdCrTiO$_5$ has been investigated via X-rays and neutron diffraction in 1970.\cite{neutron} The spins of the Cr and Nd ions were found to be antiferromagnetically (AFM) ordered below 13 K with all magnetic ions participating in the magnetic order. The spins of the Cr ions are collinearly oriented along the $c$-axis, while the Nd moments are ordered in the $ab$-plane. Subsequently, the ME effect in NdCrTiO$_5$ was discovered.\cite{greenblatt} NdCrTiO$_5$ was one of the first ME materials possessing two distinct magnetic sublattices, namely Cr$^{3+}$ and Nd$^{3+}$, known at that time. The ME susceptibility was measured \cite{greenblatt} on a powder pellet sample showing a $T_N$ of 20.5 K, rather than 13 K.\cite{neutron} The magnetic point symmetry (mmm$^\prime$) allows a linear ME effect.\cite{itc} Hence, under an applied electric field a magnetization is induced, $M_i$ = $\alpha_{ij} E_j$, where $\alpha_{ij}$ showed a rapid increase at $T_N = 20.5$ K and a drop below 8 K.\cite{greenblatt} The magnitude of $\alpha_{ij}$ observed was of the order of 1 $\times 10^{-5}$ CGS units, i.e. smaller by one order of magnitude when compared to the well known ME material Cr$_2$O$_3$.\cite{Rado} In Ref. \onlinecite{greenblatt} it is argued that below $T_N$ the magnetic moments of the Cr$^{3+}$ ions order cooperatively and then induce the order of the Nd$^{3+}$ ions by means of an exchange coupling, in contrast to a direct cooperative ordering of the Nd$^{3+}$ spin subsystem.

Assuming that the AFM order in NdCrTiO$_5$ is driven by the Cr$^{3+}$ spins, their collinear order would suggest that the mechanism giving rise to the ME is magnetostriction. However, if both magnetic sublattices  participate in the built-up of the cooperative AFM order, then the order has to be considered noncollinear and the spin-current model would be applicable. The open question of the origin of the coexistence of AFM and the ME effect motivated us to present a detailed study of the ME behavior of this compound.

In this work, we measure the electric response (dielectric constant, polarization) of NdCrTiO$_5$ under magnetic field, which leads to observations of signatures of multiferroic and ME behavior such as a dielectric anomaly and spontaneous switchable polarization below $T_N$. For the measured magnetic field range up to 7 T, the saturated polarization ($P_{sat}$) increases with magnetic fields with a maximum $P_{sat} \sim 3.5~\mu$C/m$^2$. The observed ME annealing effect tends to point NdCrTiO$_5$ as ME material rather than multiferroic, but we cannot rule out the possibility that it is a multiferroic.

\section{Experimental}

Polycrystalline NdCrTiO$_{5}$ was made by solid state reaction. A stoichiometric mixture of Nd$_{2}$O$_{3}$, Cr$_{2}$O$_{3}$, and TiO$_2$ was ground together and calcinated in air at 1000 $^{\circ}$C for 24 hours. The powder was then reground and pressed into pellets under 400 atm hydrostatic pressure, and again calcinated in air at 1300 $^{\circ}$C for another 24 hours. The x-ray powder diffraction (XRD) patterns were recorded by a HUBER Imaging Plate Guinier Camera 670 with Cu $K_{\alpha1}$ radiation (1.54059 {\AA}) with a Ge monochromator. The DC magnetic susceptibility measurements were made with a Quantum Design superconducting interference device (SQUID) magnetometer with applied magnetic field of $H = 0.1$ T. The specific heat measurements were performed in a PPMS (Physical Property Measurement System by Quantum Design). A sample of cylindrical shape was cut from the pressed pellet with dimensions of 16.0 mm$^2$$\times$1.0 mm for the measurements of the electrical properties and electrodes were painted with silver paste on the plane surfaces. The dielectric constant was measured using an Andeen-Hagerling AH-2700A capacitance bridge operating at a frequency of 10 kHz. The pyroelectric current was measured using a Keithley 6517A electrometer on warming after poling the crystal in an electric field while cooling down from above $T_{\text{N}}$. The sample was short-circuited for 15 minutes before warming up to remove any residual charges. The spontaneous polarization was obtained by integration of the pyroelectric current with respect to time.

\section{Results}

\begin{figure}[tbp]
\linespread{1}
\par
\includegraphics[width=3in]{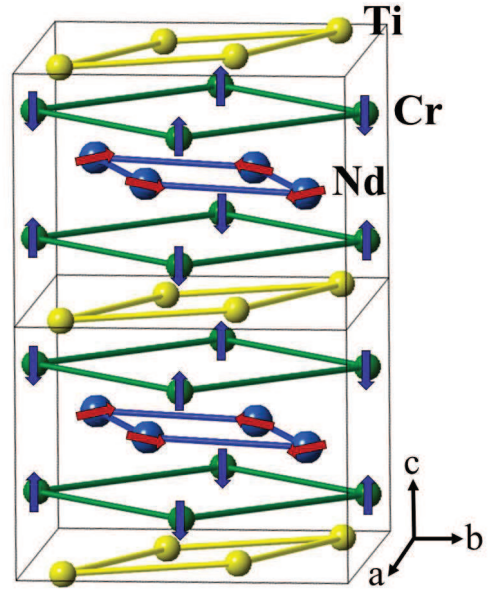}
\par
\caption{(Color online) The crystallographic and magnetic structure of NdCrTiO$_5$ according to Ref. \onlinecite{neutron}.  Here we ignored the small interchange of sites found for Cr and Ti atoms (see text).}
\end{figure}

\begin{figure}[tbp]
\linespread{1}
\par
\includegraphics[width=3.2in]{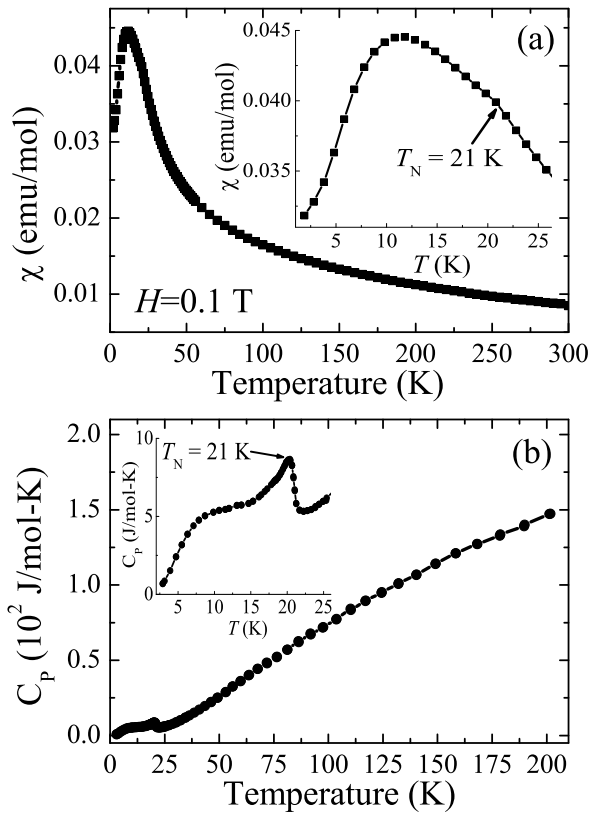}
\par
\caption{Temperature dependence of (a) the magnetic susceptibility and (b) the specific heat of NdCrTiO$_5$. Each inset shows data over an extended temperature range around $T_{\text{N}}$. Note the kink in the susceptibility at 21 K.}
\end{figure}

The room-temperature X-ray diffraction (XRD) pattern of  NdCrTiO$_5$ shows a single phase with an orthorhombic, $Pbam$, structure. The lattice parameters are $a$ $=$ 7.5812(2) {\AA}, $b$ $=$ 8.6803(2) {\AA} and $c$ $=$ 5.8123(3) {\AA}, similar to the previously reported data.\cite{neutron} The X-ray and neutron diffraction study \cite{neutron} identified that 95\% of Cr$^{3+}$ and 5\% Ti$^{4+}$ are distributed over the 4f sites (center of oxygen coordinated octahedral) and 5\%Cr$^{3+}$ - 95\% Ti$^{4+}$ over the 4h sites (base center of a square pyramid) while Nd$^{3+}$ ions are located at the 4g sites.

The DC magnetic susceptibility (see Fig. 2(a)) shows a slope change (kink) at 21 K. At the same temperature, the specific heat shows a $\lambda$ shape peak, which indicates that this transition involves long range magnetic order. This transition temperature is in good agreement with that previously reported from ME measurements.\cite{greenblatt} The earlier reported neutron diffraction study \cite{neutron} has identified the low temperature phase ($T \le 13$ K) as having AFM order. The magnetic moments of Cr$^{3+}$ in the 4f sites (Cr (4f)) are pointing along the $c$-axis, with alternating up-ward and down-ward orientations. The magnetic moments of Nd$^{3+}$ and Cr$^{3+}$ in the 4h sites (Cr (4h)) lie in the $ab$ plane. The Cr (4h) moments are aligned along the $b$-axis and AFM correlated in that direction, while they are ferromagnetically correlated in the $a$-direction. The Nd moments, on the other hand, order in a similar way except that they are tilted away from the $b$-axis by 12 $\deg$. The long range order of the Nd moments is believed to follow the ordered Cr moments rather than being induced by a cooperative interaction of Nd$^{3+}$ spin systems.\cite{greenblatt,yaeger} The spin structure at temperatures well below $T_{\text{N}}$ (ignoring the 5\% of switched Ti and Cr ions between the 4h and 4f sites) and below the temperature of the magnetic susceptibility peak ($T$ = 11 K) is depicted in Fig. 1.  Note that the slope of the susceptibility begins to change below 13 K, which is reported in Ref. \onlinecite{neutron} as the temperature below which the magnetic Bragg peaks were observed.

\begin{figure}[tbp]
\linespread{1}
\par
\includegraphics[width=3.5in]{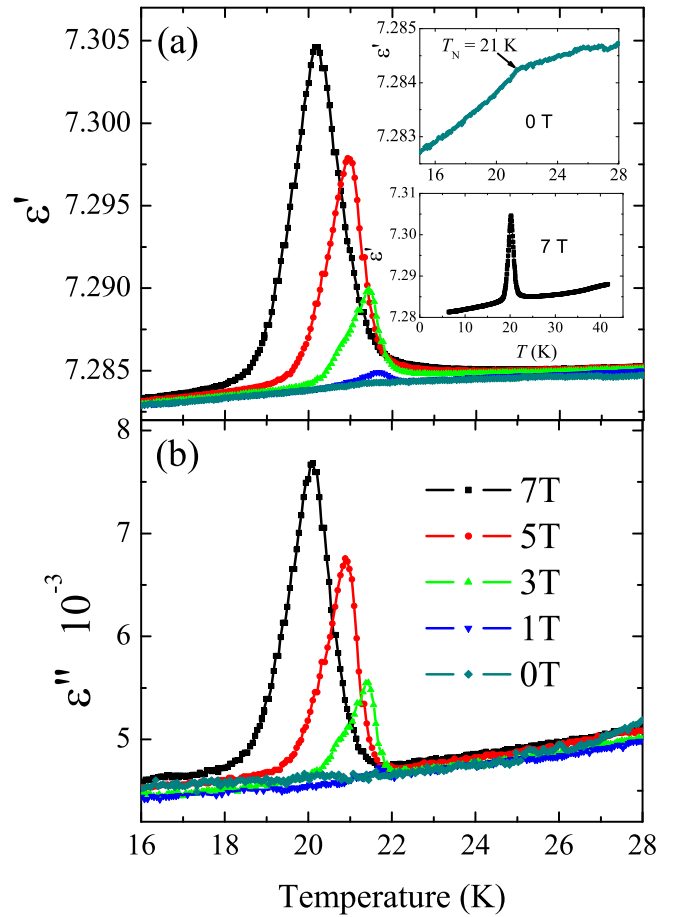}
\par
\caption{(Color online) The temperature dependence of the real part (a) and the imaginary part (b) of the dielectric constant of NdCrTiO$_5$ at different applied magnetic fields. The upper inset of (a) shows the dielectric constant with zero magnetic field over an extended temperature range around $T_{\text{N}}$. Note the kink at the antiferromagnetic transition. The lower inset of (a) shows the dielectric constant down to 6.5 K at 7 T.}
\end{figure}

Fig. 3 shows the dielectric constant measured for different applied magnetic fields. For $H$ = 0 T, the real part of the dielectric constant ($\varepsilon'$) shows a slope change around $T_{\text{N}}$ = 21 K. When a magnetic field is applied, a peak is induced in $\varepsilon'$ as shown in Fig. 3(a). Since $T_{\text{N}}$ decreases with magnetic field in an antiferromagnet (the order is gradually suppressed), with increasing magnetic field, the position of the peak moves to lower temperatures. Since the spin symmetry is broken by the applied magnetic field, the peak becomes stronger with increasing field. We also notice that (i) the position of the peak in $\varepsilon'$ shows no frequency dependence (not shown here); (ii) the imaginary part of the dielectric constant, or the dissipation, also manifests a peak around a similar temperature as the peak of $\varepsilon'$ with applied magnetic field. These two behaviors prove that the anomaly of $\varepsilon'$ is really related to the phase transition. Hence, the observation of the anomaly of $\varepsilon'$ around the AFM ordering transition at $T_{\text{N}}$ = 21 K demonstrates the coupling between the magnetic and electric properties of NdCrTiO$_5$. We did not observe any anomaly associated with the broad peak in the magnetic susceptibility around $T$= 11 K.

\begin{figure}[tbp]
\linespread{1}
\par
\includegraphics[width=3.2in]{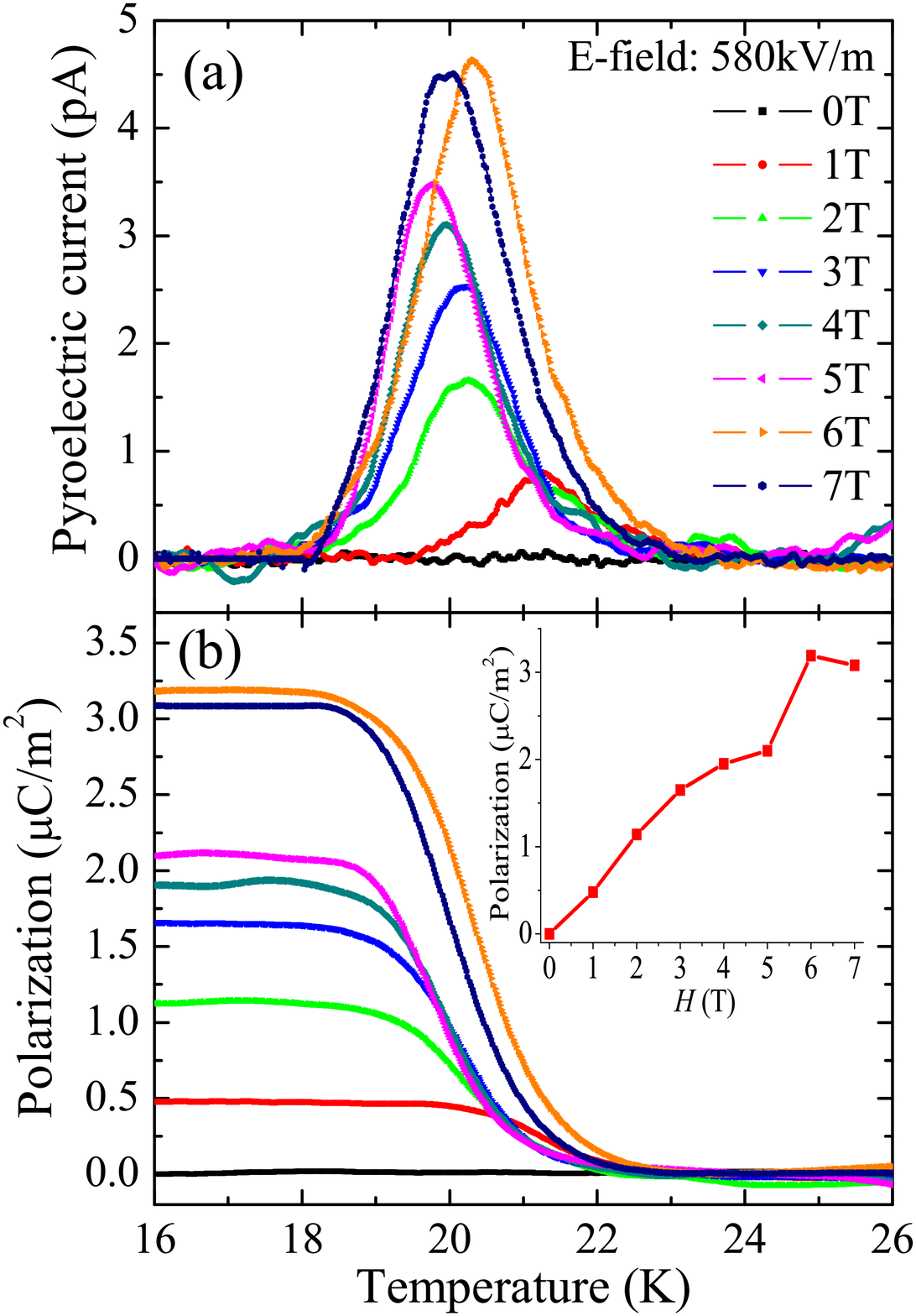}
\par
\caption{(Color online) Temperature dependence of the pyroelectric current (a) and polarization (b) of NdCrTiO$_5$ for different applied magnetic fields for a fixed electric field of 580 kV/m (E$\bot$H). The inset of (b) shows the magnetic field dependence of the low $T$ polarization.}
\end{figure}

Fig. 4(a) shows the pyroelectric current measured for different magnetic fields and a fixed electric field. The magnetic fields were applied during both cooling and warming up. In zero magnetic field, no pyroelectric current is observed. When a magnetic field is applied, a pyroelectric current arises in a temperature interval related to the phase transition. This broad peak of pyroelectric current starts at about 18 K and ends around 22 K and becomes stronger with increasing applied magnetic field.  Note that it is necessary to cool the sample through $T_{\text{N}}$ with an applied electric field to observe a pyroelectric current. The spontaneous polarization is obtained by integration of the pyroelectric current with respect to time and shown in Fig. 4(b). With increasing applied magnetic field, the polarization monotonically increases and saturates around 3.2 $\mu$C/m for $H$ around 6 $\sim$ 7 T. The pyroelectric current and the onset of polarization are then directly related to the magnetic phase transition. Note that again no change in the pyroelectric current was observed around $T$= 11 K, where the magnetic susceptibility has its maximum.

\begin{figure}[tbp]
\linespread{1}
\par
\includegraphics[width=3.2in]{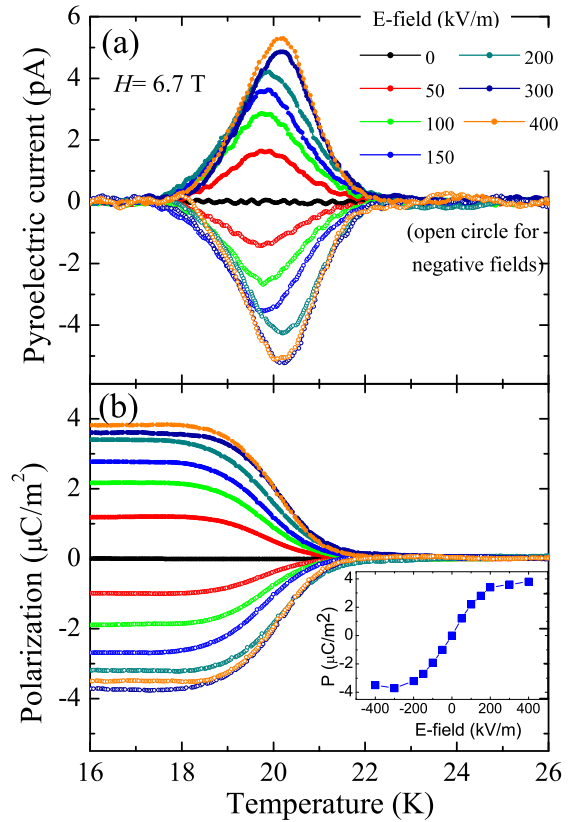}
\par
\caption{(Color online) Temperature dependence of the pyroelectric current (a) and polarization (b) of NdCrTiO$_5$ at different electric fields and a magnetic field $H$ of 6.7 T (E$\bot$H). The solid and open circles correspond to reversed electric fields. The inset shows the polarization as a function of poling electric fields.}
\end{figure}

Fig. 5 displays the pyroelectric current and its integrated polarization measured on a sample from a different batch for various applied electric fields under a fixed magnetic field $H$ = 6.7 T. The peak of the pyroelectric current around $T_{\text{N}}$ and its related polarization become stronger with increasing applied electric field. Moreover, the sign of the pyroelectric current and polarization switches simultaneously with the sign of the applied electric field. As shown by the solid and open circles in Fig. 5, a similar pyroelectric current of opposite sign is obtained and hence a polarization of the same magnitude. This is typical for ME annealing conditions, i.e., poling with both $H$- and $E$-fields. The polarization increases linearly with applied electric fields up to 200 kV/m followed by saturation behavior at higher electric fields.

\begin{figure}[tbp]
\linespread{1}
\par
\includegraphics[width=3.5in]{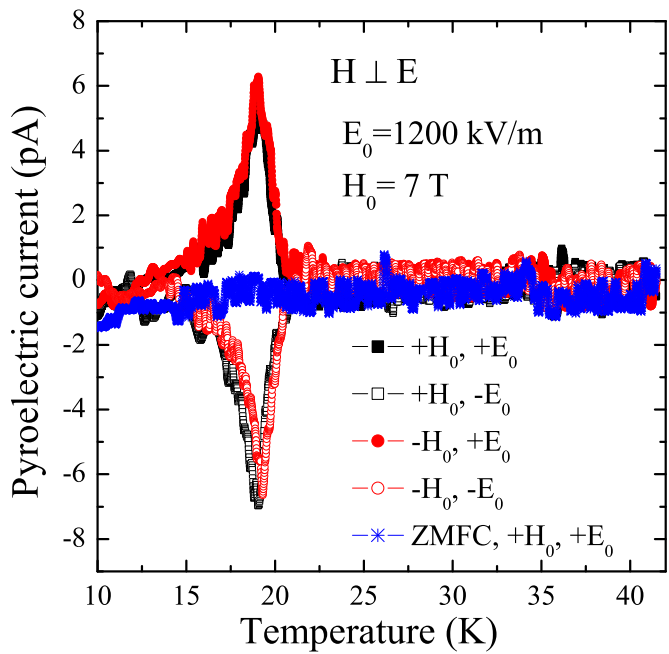}
\par
\caption{(Color online) Temperature dependence of the pyroelectric current NdCrTiO$_5$ under different ME annealing conditions when the electric field is perpendicular to the magnetic field (E$\bot$H).}
\end{figure}

We studied the ME annealing effect more closely on another sample from the same batch measured in Fig. 5. For this study, we empolyed five different condition of ME annealings, namely, four different conditions of positive or negative combination of magnetic or electric fields and zero magnetic field cooling (ZMFC). As stated above, to observe polarization, the electric field is always applied during the annealing (cooling) process, and so is the magnetic field during the pyrolelectric current measurement (warming up). Fig. 6 shows the pyroelectric current data from this study when the electric field is applied perpendicular to the magnetic field. From this study, we found that the usual ME annealing condition (both magnetic and electric fields applied during cooling) is necessary to induce polarization. The polarization reversal is achieved by changing the polarity of the electric field but not the magnetic field.
The same measurements were repeated after the sample was rotated by 90 degree so that the electric and the magnetic fields are parallel each other. The two different configurations (E$\bot$H or E $\|$H) gave same results but with 50\% larger polarization value for the E$\bot$H case.

\section{Discussion}

The maximum spontaneous polarization value (3.5 $\mu$C/m) is smaller by two orders of magnitude as compared to other well-known multiferroic materials with the same structure, e.g., RMn$_2$O$_5$ (R: rare earth elements or Y). The polarization value is expected to be larger in NdCrTiO$_5$ when measured for a single crystal, rather than for a polycrystalline sample, but is likely to remain within the same order of magnitude. Considering the almost linear increase of the polarization with magnetic fields (see the inset of Fig. 4(b)), larger magnetic fields could also enhance the polarization. From the linear relation, we obtained ME coefficients $\alpha_{xy}$ = 1.5 $\times$ 10$^{-5}$ CGS unit and $\alpha_{zz}$ = 0.7 $\times$ 10$^{-5}$ CGS units. These values are in the same order with what was obtained from the ME susceptibility measurement.\cite{greenblatt}

Our findings, the dielectric constant anomaly and the polarization reversal, suggest two possible scenarios; (1) multiferroicity with a concurrent AFM/FE transition or (2) a linear ME effect with AFM fluctuation and domain effects. A high resolution structural measurement or a polarization-electric field ($PE$) hysteresis experiment at low temperatures could discern the two scenarios, but neither of them is available to us at the moment. At room temperature, NdCrTiO$_5$ is orthorhombic (Pbam), a centrosymmetric structure at room temperature, which does not allow ferroelectricity via the DM interaction. Nevertheless, we cannot exclude the AFM/FE scenario  without further definitive experiments. Hereafter, we discuss each scenario and its implications.

\subsection {Multiferroicity with concurrent AFM/FE transition}

In this case, the mechanism leading to FE is associated with the AFM phase transition. Due to the rather small magnitude of the polarization, the corresponding driving force inducing the symmetry breaking must be weak. To find the underlying mechanism, we examine the applicability of the spin current model first. The origin of the downturn of the magnetic susceptibility below 11 K is not yet established. A possible plausible scenario is that the Nd$^{3+}$ spins (the $J=9/2$ multiplet is split by crystalline fields to yield a Kramers ground doublet) gradually start to participate in the AFM order below 13 K, a temperature reported in Ref. \onlinecite{neutron} as the onset of some of the magnetic Bragg reflections. The onset of the order of the Cr$^{3+}$ moments could not be determined in Ref. \onlinecite{neutron} and is seen as a very weak cusp in the magnetic susceptibility (see inset in Fig. 2(a)). At $T_N$ the fluctuations of the magnetic moments are rather weak compared to the thermal fluctuations at the transition. This picture agrees with the arguments presented in the earlier papers about this compound.\cite{neutron,greenblatt} Assuming that close to $T_N$ only the Cr$^{3+}$ moments order, then the magnetic structure is collinear, so that ${\vec S_i} \times {\vec S_j}$ terms are zero in the temperature interval around $T_N$. The spin current mechanism\cite{Katsura,Sergienko} can then be excluded, leaving magnetostriction\cite{goodenough} as the candidate to induce the FE.

The above arguments are based on the assumption that the Nd$^{3+}$ and the Cr$^{3+}$ (4h) moments are not participating in the magnetic order until lower temperatures. Should this not be the case, then spin cross-products between Cr (4f) and Nd sites are nonzero and since the system lacks inversion symmetry, the DM mechanism is active to produce a spin current. The strength of the spin current would depend on the magnitude of the magnetic order in the Nd sublattice. Such mechanism, however, seems to be absent in NdCrTiO$_5$ considering that no features of the dielectric constant and the polarization were observed around the Nd ordering temperature.

In order to explain the dielectric anomaly at zero field (see the inset of Fig. 2(a)) and the almost linear magnetic field dependence of the polarization we turn to the Ginzburg-Landau expansion of the free energy in terms of the electric and magnetic fields.\cite{Fiebig} Differentiation with respect to the fields leads to the polarization and the magnetization,
\begin{eqnarray}
P_i({\vec E},{\vec H}) &=& P_i^S + \epsilon_0 \epsilon_{ij} E_j + \alpha_{ij} H_j + \cdots \ , \\
M_i({\vec E},{\vec H}) &=& M_i^S + \mu_0 \mu_{ij} H_j + \alpha_{ij} E_j + \cdots \ ,
\end{eqnarray}
where ${\vec P}^S$ and ${\vec M}^S$ denote the spontaneous polarization and magnetization, whereas ${\hat \epsilon}$ and ${\hat \mu}$ are the electric and magnetic susceptibilities. The tensor ${\hat \alpha}$ represents the induction of polarization by a magnetic field or of magnetization by an electric field which is known as the linear ME effect. For an AFM we have ${\vec P}^S = {\vec M}^S =0$, so that in zero magnetic field an electric field can induce a finite magnetization, $M_i$ = $\alpha_{ij} E_j$, and vice versa, in zero electric field a magnetic field can induce a finite polarization, $P_i$ = $\alpha_{ij} H_j$. We have mentioned this effect in the Introduction.

In Fig. 4(b) we presented the polarization as a function of temperature for several fields. The magnetostriction, when the sample is driven through the N\'eel transition, induces strains and hence a local magnetization in the grains and their boundaries. The small displacements of the ions from their equilibrium position, the consequence of the magnetostriction, should change the polarizability of the sample and hence the dielectric constant. Also the local deviations from the ordered magnetization, $\delta m_i$, arising from the magnetostriction should generate local electric fields through the ${\hat \alpha}$-tensor, giving rise to a change in the dielectric constant at $T_N$. This effect is expected to be very weak in the absence of a magnetic field, but it is very magnetic field sensitive since all the local contributions $\delta m_i$ align and become cooperative. This effect is restricted to the region around $T_N$. It is also expected that the zero magnetic field anomaly of $\epsilon'$ is stronger in a polycrystalline sample, due to the numerous grain boundaries, than in a single crystal. Its magnetic field dependence should, however, be stronger in a single crystal. The effect on the absorption, i.e. $\epsilon''$, is expected to be much weaker than for $\epsilon'$ as indeed observed.

The pyroelectric current generated when the system is cooled through the N\'eel transition is also expected to be proportional to $\delta m_i E_j$. Hence, it is restricted to the region around $T_N$, grows with applied magnetic field approximately linearly and is linear in the electric field. The saturation polarization is then almost linear in the applied $H$, in agreement with the observations. The above is valid as long as the expression $P_i$ = $\alpha_{ij} H_j$ is valid, i.e. the fields are not too large; otherwise higher order terms in the Ginzburg-Landau expansion would have to be taken into account.

A specific mechanism of how the magnetostriction exchange can lead to a broken inversion symmetry in NdCrTiO$_5$ below the N\'eel transition is not available yet. In other multiferroic materials with collinear spin structures, the interactions between neighboring spins from different sites seem to play a crucial role. For example, there are ferromagnetically aligned Dy$^{3+}$(Gd$^{3+}$) and Fe$^{3+}$ spin sheets in Dy(Gd)FeO$_3$\cite{Tokunaga,gdfeo3} and $\uparrow \uparrow \downarrow \downarrow$ order of Co$^{2+}$ and Mn$^{4+}$ Ising spins in Ca$_3$Co$_{2-x}$Mn$_x$O$_6$.\cite{choi} This kind of spin arrangement is not clearly apparent for NdCrTiO$_5$ according to the available spin structure (see Fig. 1). Similar measurements on a single crystal sample would give more insight on the preferred polarization direction and its magnetic field dependence.

\subsection {Non-FE transition/Linear ME effect}
First of all, it should be noted that the linear ME effect and the multiferroicity are not mutually exclusive, that is, they can occur in a single material as found in DyFeO$_3$.\cite{Tokunaga} Experimentally, it is easier to observe the linear ME effect compared to proving ferroelectricity. We attempted $PE$ hysteresis measurement on NdCrTiO$_5$ samples but the experiment was inconclusive since we could not observe typical hysteresis behavior before the sample was damaged due to electrical breakdown.

There are characteristic behaviors observed in NdCrTiO$_5$ which favor the non-FE/linear ME effect scenario. For example, the ME annealing behavior shown in Fig. 6 is typical for powder ME material. It is well known that the ME annealing is necessary to observe ME effect in AFM material, otherwise random AFM domains result in zero net polarization.\cite{hornreich, Rado2, Shtrikman} The reversal of the polarization was found in other ME materials when the ME annealing was done with opposite polarity of electric field.\cite{Nenert}

Likewise, other experimental results can be also explained without evoking ferroelectricity; e.g. the linear field dependent polarization and the larger ME coefficient for (E$\bot$H). The observed dielectric anomaly around $T_N$ is often found in other AFM systems, where spin fluctuations are ascribed to the anomaly.\cite{Lawes2}. The recovery of the dielectric constant well below the transition and the increased anomaly with the magnetic field can as well be qualitatively explained by this model.

However we can not completely exclude the possibility of multiferroicity. For example, it is also conceivable that the ME annealing is required for magnetic field induced FE transition. The poling process is to acquire a single domain state while the sample is cooled down through a transition (for example, paraelectric to ferroelectric), therefore both magnetic and electric fields are necessary under the AFM/FE scenario as well.
Between the two scenarios, however, we tend to incline to the second one, the linear ME effect rather than multiferroics. It is the simpler explanation and it is consistent with the small polarization.

In summary, we have observed a field induced dielectric constant anomaly and electric polarization at the AFM state of NdCrTiO$_5$. The polarization increases with magnetic fields and is reversible by changing the polarity of the poling electric field. We discussed multiferroicity and linear ME effect as a possible mechanism to explain the experimental results. While it is difficult to draw a definitive conclusion, the linear ME effect is more likely the cause considering the small polarization.

\begin{acknowledgments}
This work was performed at the National High Magnetic Field Laboratory which is supported by NSF Cooperative Agreement No. DMR-0654118 and by the State of Florida. P.S. is supported by the DOE under Grant No. DE-FG02-98ER45707.
\end{acknowledgments}

\end{document}